\documentclass[a4paper]{jpconf}
\usepackage{graphicx}
\usepackage{hyperref}
\usepackage{gensymb}
\usepackage{bm}
\usepackage{subcaption}
\usepackage{caption}

\begin{document}

\title{Recent design studies for the novel momentum spectrometer NoMoS}

\author{D Moser$^1$, W Khalid$^1$, R Jiglau$^1$, T Soldner$^2$, M Valentan$^1$, \mbox{J Zmeskal$^1$} and G Konrad$^1$}

\address{$^1$ Stefan Meyer Institute, Austrian Academy of Sciences, Boltzmanngasse 3, 1090 Wien, Austria}
\address{$^2$ Institut Laue-Langevin, 71 avenue des Martyrs
CS 20156, 38042 Grenoble Cedex 9, France}

\ead{daniel.moser@oeaw.ac.at}

\begin{abstract}
NoMoS is a novel momentum spectrometer with which we aim to measure the spectra of the charged neutron beta decay products with high precision. The shape of the proton and electron spectra can inter alia be used for the determination of the electron-antineutrino correlation coefficient $a$ and the Fierz interference term $b$, respectively. These observables can in turn be used to test the Standard Model of Particle Physics and to search for extensions thereof. NoMoS utilizes the $R\times B$ drift effect present in curved magnetic fields, which disperses charged particles according to their momentum. In this paper, we report on selected recent investigations that were conducted with regard to the magnet design and the detection system. 
\end{abstract}

\section{Introduction}

NoMoS, the \underline{N}eutron decay pr\underline{o}ducts \underline{Mo}mentum \underline{S}pectrometer, is designed to measure momentum spectra primarily of the charged decay products in free neutron beta decay \cite{Konrad15, Mos19}. Measuring the spectra of the electrons and protons opens the door to several interesting observables, such as the electron-antineutrino correlation coefficient $a$ or the Fierz interference term $b$. They affect the shape of the proton and electron spectrum, respectively. A precise determination of these observables is a strong tool to test the Standard Model of Particle Physics (SM) and to search for new physics beyond it \cite{Profumo07, Dubbers2011, Cirigliano2013, Marciano18, GONZALEZ19}. 

Numerous experiments have been built or are under construction to measure these observables with high precision: conducted measurements of $a$ have achieved an accuracy of $\Delta a/a=2.6$\% \cite{PDG18} (PDG) whereas $b$ was determined with an accuracy of $\Delta b = 0.09$ \cite{Hickerson17}. Currently, various experiments try to improve on this. aCORN is still in the process of data taking with an ultimate goal of $\Delta a/a \sim 1$\% \cite{Darius17}, \textit{a}SPECT recently published its final value with an accuracy of $\Delta a/a=0.8$\% \cite{Beck2019} and Nab aims for $\Delta a/a \sim 0.1$\% \cite{Fry18}. Nab also plans to measure the Fierz interference term with an ultimate precision of $\Delta b < 10^{-3}$ \cite{Broussard2019}. The ultimate goal of NoMoS is $\Delta a/a < 0.3$\% and $\Delta b<10^{-3}$.

In this paper, we present handpicked in-depth investigations of the apparatus design. These include a comparison of two magnet configurations and arising changes in systematic effects as well as the systematic effect of detector misalignment.

\section{Experimental Design of NoMoS}

\begin{figure}[h]
\centering
\includegraphics[width=0.48\textwidth]{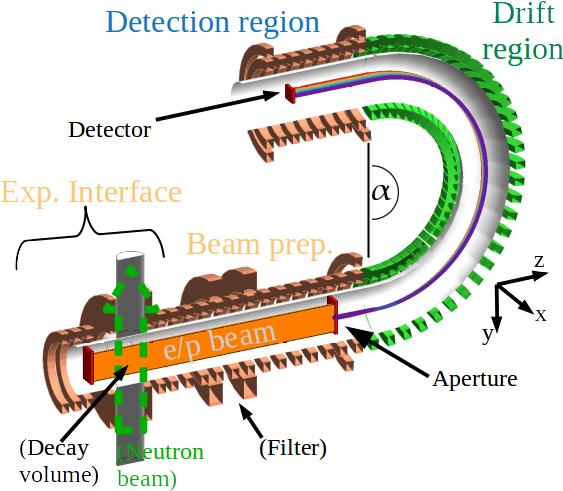}\hspace{1pc}%
\begin{minipage}[b]{0.48\textwidth}\caption{\label{fig:RxB}\textbf{Scheme of NoMoS:} The apparatus is divided into four areas: the experimental interface, the beam preparation area, the drift, and the detector regions. Optional features for in-situ particle collection are drawn dashed or labeled in parentheses. \textbf{Particle beam:} The electron/proton beam is magnetically guided through the system and geometrically shaped by the aperture. In the drift region the particles drift according to their momentum. The drift distance is measured with a detector in the detection region.}
\end{minipage}
\end{figure}

NoMoS is an $R \times B$ drift momentum spectrometer, getting its name from the so-called $R \times B$ drift effect. This drift appears in curved magnetic fields (derived from the gradient drift) and the drift velocity is in zeroth order proportional to \cite{JacksonEdyn}:

\begin{equation}
\vec{v}_{R\times B} \propto \frac{\vec{R} \times \vec{B}}{qR^2B^2}
\end{equation}
\noindent
with the curvature vector $\vec{R}$ and the magnetic flux density $\vec{B}$ as well as their respective vector norms $R$ and $B$. As the drift velocity also depends on the charge $q$ of the particle, electrons and protons drift in opposite directions. When the velocity is integrated over time, the resulting drift distance is linearly proportional to the particle's momentum $p$ \cite{Wang2013}:

\begin{equation}
D_{R\times B} = \frac{p \alpha}{qB} f(\theta)
\label{equ:drift}
\end{equation}
with $\alpha$ being the curvature angle of the curved field region (see Fig.~\ref{fig:RxB}) and \mbox{$f(\theta)=\left(\cos\theta+1/\cos\theta\right)/2$} describing the dependence on the particle's incident angle $\theta$.

The principal set-up of NoMoS is shown in Fig.~\ref{fig:RxB}. It consists of the experimental interface, the beam preparation area, the drift region as well as the detection region. First, the charged neutron decay products have to be magnetically collected in a decay volume and magnetically guided towards the rest of the apparatus. This can either be realized in a facility like PERC \cite{Dubbers08, PERC12} (which also has its own magnetic filter for precise incident angle selection) or in-situ in the experimental interface. Then, the particle beam is further manipulated in the beam preparation area. An optional in-situ magnetic filter can be installed, that repels particles with an incident angle larger than a critical angle. An aperture defines the beam cross-section before the particles pass through the curved drift region, in which they experience the $R \times B$ drift. A spatially resolving detector is placed in the detection region to measure the particles' positions from which the drift distances can be inferred.

\section{Investigation of an Alternative Magnet Configuration}
The magnet system of NoMoS consists of several parts that significantly impact the transport of the particles through the apparatus, like the magnetic filter. These parts' particular realization influences the functionality of NoMoS as well as the sensitivity of the observables on several systematic effects. Therefore, different magnet design configurations are reviewed for advantages and disadvantages. 

One interesting variation of the NoMoS magnet (compared to the design presented in \cite{Mos19} and Fig.~\ref{fig:RxB}) is the combination of the following two changes:

\begin{itemize}
\item \textit{Curvature angle $\alpha$:} only 90\degree \,of curvature angle in the drift region instead of 180\degree , and
\item \textit{Incident angle manipulation:} Lowering the incident angles with a decreasing B-field (inverse magnetic mirror effect) instead of selecting through the magnetic mirror effect with a B-field maximum (filter).
\end{itemize}
\noindent
Not only does this configuration extensively modify the system's geometry, it also changes a number of systematic effects. This is predominantly because $\alpha$ strongly influences the overall shape of the drift distance spectrum (see linear dependence in Equ. \ref{equ:drift}). %Moreover, the sensitivity of incident angle effects is differently distributed between the single B-field parameters. 

\begin{figure}[h]
\centering
\begin{minipage}[l]{0.48\textwidth}
\includegraphics[trim={0 0 0 1.5cm}, width=\textwidth]{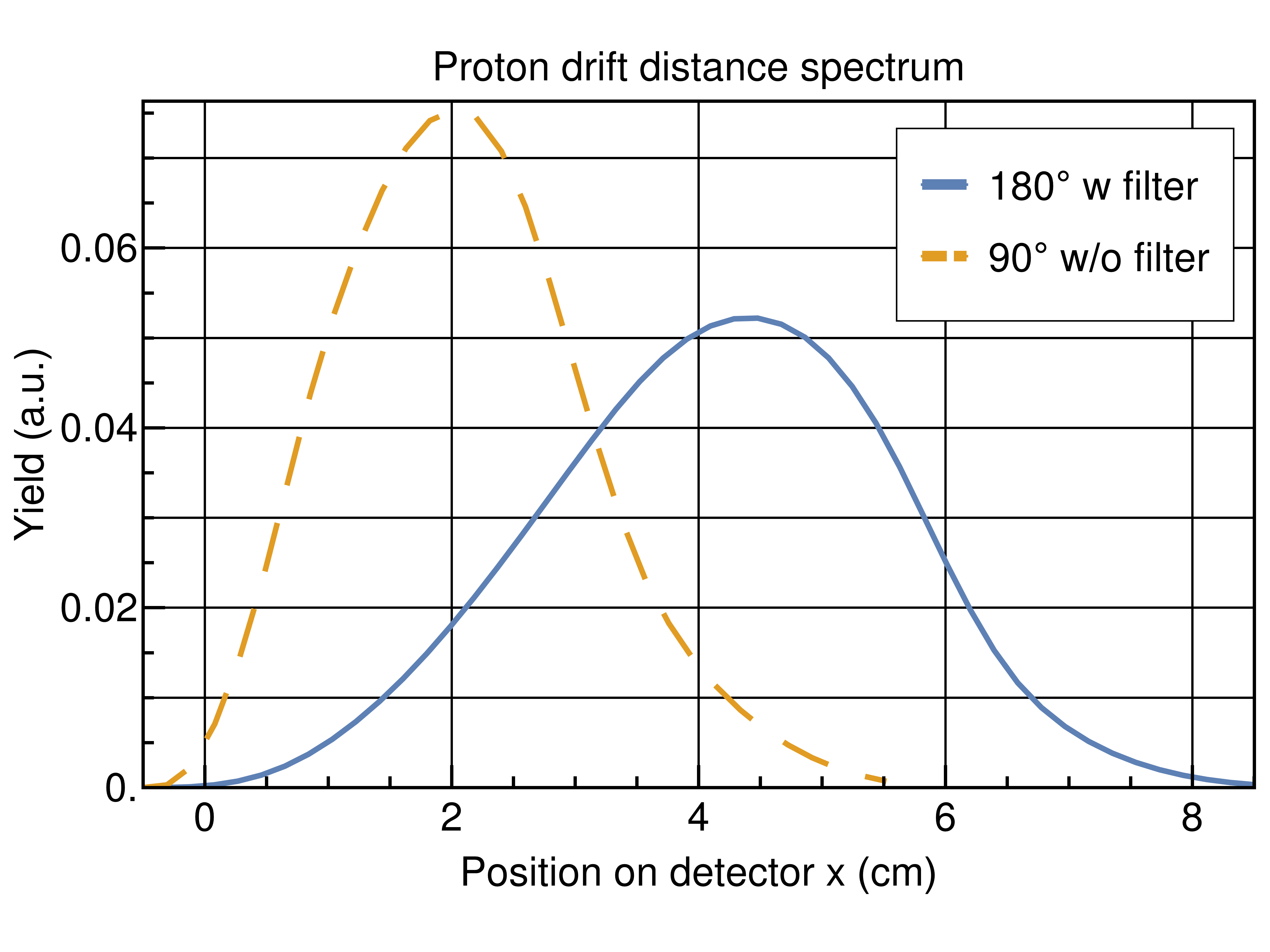}
\subcaption{\label{fig:transporta}A proton drift distance spectrum comparing the two different configurations.}
\end{minipage}
\hspace{1pc}%
\begin{minipage}[r]{0.48\textwidth}
\includegraphics[width=\textwidth]{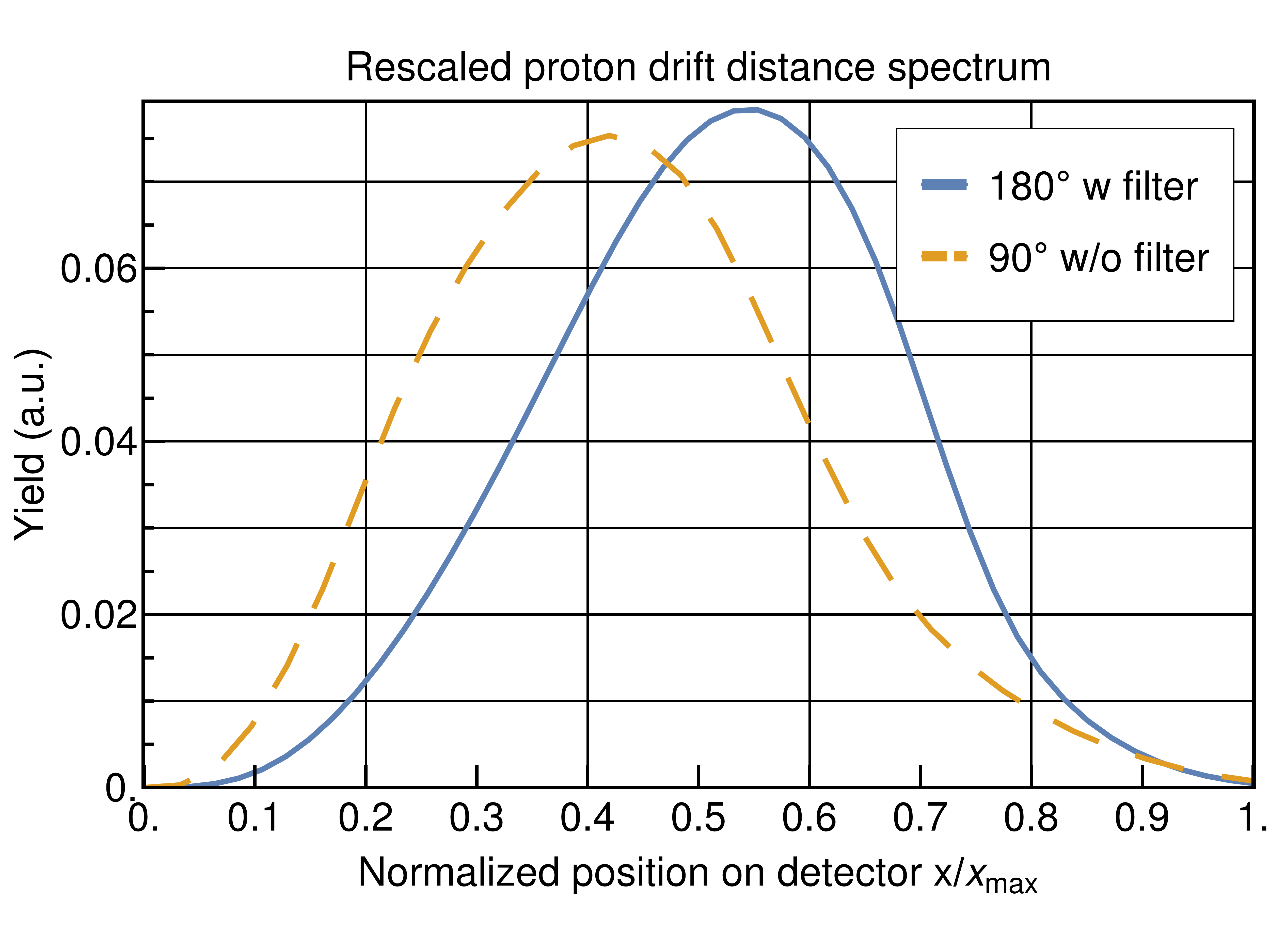}
\subcaption{\label{fig:transportb}The proton drift distance spectra of both configurations with normalized x-axis.}
\end{minipage}
\caption{\label{fig:transport}An examplatory proton drift distance spectrum, comparing $\alpha=180\degree $\, with filter (blue, solid) to $\alpha=90\degree \,$ without filter (orange, dashed). Other parameters of this spectrum: Absolute B-field level in the drift region $B_{R \times B}= 0.2$ T, $a=-0.103$, the aperture opening in drift direction $x_A = 1$ cm, the ratios of B-fields in specific areas relative to the decay volume $r_{A,R \times B,Det}= 1$ and $1/2$, $r_F= 2$ and $1$ (aperture, drift region, detector, filter), for $\alpha=180\degree$ and $90\degree$, respectively.}
\end{figure}

Figure \ref{fig:transport} shows an examplatory proton drift distance spectrum for $\alpha=180\degree$ with filter compared to $\alpha=90\degree$ without filter. Figure \ref{fig:transporta} shows the strong effect of $\alpha$ shifting the spectrum to lower drift distances for 90\degree , whereas Fig.~\ref{fig:transportb} illustrates the change in shape with a rescaled spectrum. In the latter, we observe a more asymmetric and overall broader curve for the 90\degree \,configuration.

We investigate the sensitivity of the observables on systematic parameters with the help of a fit function, that includes all particle transport effects and the observables through the system. This function is fitted to drift distance spectra with minor changes introduced in the systematic effects' parameters. This mimics a small error in the observables resulting from an uncertainty in the systematic parameters.

Two significant systematic effects in both configurations are from uncertainties in $B_{R \times B}$ and $\alpha$. In the following, we assume an uncertainty in the parameters of $\Delta B_{R \times B} / B_{R \times B} = 1\times 10^{-4}$ and $\Delta \alpha / \alpha = 4\times 10^{-4}$. In Table \ref{tab:sys} we list preliminary errors in $a$ and $b$ from those systematic effects.

\begin{center}
\begin{table}[h]
\caption{\label{tab:sys}Preliminary systematic errors in $a$ and $b$ from uncertainties in $B_{R \times B}$ and $\alpha$ determined by a fit function (as described above) for the two aforementioned magnet configurations. For both systematic effects, an additional fit parameter is introduced. All errors given are upper bounds. The values are given for the in-situ particle collection (PERC would introduce its own filter - this in general results in smaller systematic effects). All systematic parameters' values are given in Fig.~\ref{fig:transport}.}
\centering
\begin{tabular}{lp{2cm}p{2cm}|p{2cm}p{2cm}}
\br %bold rule
& \multicolumn{4}{c}{Systematic error ($\times 10^{-3}$)} \\
 & \multicolumn{2}{p{4cm}|}{\centering \textbf{$\bm{\alpha=}$180\degree\,w filter}} & \multicolumn{2}{p{4cm}}{\centering \textbf{$\bm{\alpha=}$90\degree \,w/o filter}}\\
 Systematic effect & $\Delta a/a$ & $\Delta b$ & $\Delta a/a$ & $\Delta b$ \\
\mr
Absolute B-field level $B_{R\times B}$& $<0.8$ & $<0.2$ &$<0.7$ &$<0.2$\\
Curvature angle $\alpha$ & $<1.8$&$<0.4$ &$<2.6$ &$<0.7$ \\
%Filter ratio $r_{F}$ &0.08 &0.04 &N.A. &N.A. \\
%Aperture ratio $r_{A}$ &calc &2.0 &1.7 &0.7 \\
%$R \times B$ ratio $r_{R \times B}$ &3.1 &0.5 &3.9 &0.9 \\
%Detector ratio $r_{Det}$ & $<0.4$* & $<0.1$* & $<1.2$* & $<0.4$* \\
\br
\end{tabular}
\end{table}
\end{center}

%The table shows that not only the systematic uncertainty due to the curvature angle $\alpha$ changes but also due to other parameters. 

Overall, the errors are worse for 90\degree\,without filter but still on a managable level (considering the upper bound nature of the estimates). In the end, besides systematic effects, also other aspects of the magnet design configuration are going to be considered for a final decision, like complexity of construction or cost.

\section{Detector Misalignment Investigations}

Since the drift distance experienced by the decay products depends on the magnetic field, the precise alignment of the spatially resolving detector to the magnetic field lines is crucial. Any mechanical inaccuracies in the alignment would introduce systematic errors in the spatial measurement of the spectrum that could dominate the desired precision for the observables.

Studies into misalignment effects reveal that an uncertainty of $\pm 0.1$\,mm in the positioning of the detector in the drift direction could generate a relative error of $>10^{-3}$ in $a$ and an absolute error of the same order in $b$. To suppress these errors, an additional fit parameter is incorporated in the fit function to account for positional uncertainties of the detector. With the help of this additional fit parameter, the systematic errors arising from potential misalignment can be suppressed by about two orders of magnitude for both observables irrespective of the magnet system configuration described in Section 3. A moderate positioning precision of $\pm 0.5$\,mm can easily be achieved using simple spatial measurement techniques, such as calliper gauges. Therefore, the fitting routine was tested with increasing misalignment distances up to $\pm 0.5$\,mm. Resulting systematic errors remain suppressed by about two orders of magnitude over the tested range.

\section{Summary and Outlook}

The purpose of this paper is to give insight into some of the numerous investigations going on within the NoMoS project to optimize the apparatus. We present an illustrative alternative magnet design, including changes in some of the systematic effects. Eventhough this configuration is still viable, we observe that the shown systematic effects get worse for the alternative magnet configuration. For a final decision on the apparatus design, also  other considerations have to be taken into account, like cost. In addition, we discuss the importance of proper alignment of the detector and cite the error a potential misalignment would introduce in the observables. We also present how we can reduce the aforementioned errors by the introduction of a fit parameter in our fit function. It is investigations like these that are used to determine the optimal design for NoMoS.

\ack
%unnumbered section acknowledgements
We would like to thank Eberhard Widmann (SMI) for his support and helpful discussions. This work is supported by the Austrian Academy of Sciences  within  the  New  Frontiers  Groups  Programme NFP 2013/09, the Austrian Science Fund under contract No.  W1252-N27 (DK-PI), the TU Wien, and the SMI Wien.

\section*{References}

% bibliography
\bibliographystyle{iopart-num}

\bibliography{references.bib}

\end{document}